\documentclass[reprint,sort&compress,aps,twocolumn,prb,superscriptaddress]{revtex4-2}

\usepackage{graphicx}
\usepackage{dcolumn}
\usepackage{color}
\usepackage{physics}
\usepackage{bm}
\usepackage{hyperref}
\usepackage{soul}
\hypersetup{
     colorlinks   = true,
     citecolor    = blue
}
\usepackage{xcolor}
\usepackage{amssymb}
\begin{document}
\newcommand{\mos}{MoS$_2$}
\preprint{APS/123-QED}


\title{Electronic properties of twisted bilayer graphene suspended and encapsulated with hexagonal boron nitride}

\author{Min Long}
\affiliation{Key Laboratory of Artificial Micro- and Nano-structures of Ministry of Education and School of Physics and Technology, Wuhan University, Wuhan 430072, China}

\author{Zhen Zhan}
\email{zhenzhanh@gmail.com}
\affiliation{Key Laboratory of Artificial Micro- and Nano-structures of Ministry of Education and School 
of Physics and Technology, Wuhan University, Wuhan 430072, China}

\author{Pierre A. Pantale\'on}
\affiliation{IMDEA Nanociencia, C/Faraday 9, 28049 Madrid, Spain}

\author{Jose \'{A}ngel Silva-Guill\'{e}n}
\affiliation{IMDEA Nanociencia, C/Faraday 9, 28049 Madrid, Spain}

\author{Francisco Guinea}
\affiliation{IMDEA Nanociencia, C/Faraday 9, 28049 Madrid, Spain}
\affiliation{Donostia International Physics Center, Paseo Manuel de Lardiz\'{a}bal 4, 20018 San Sebastián, Spain}
\affiliation{Ikerbasque. Basque Foundation for Science. 48009 Bilbao. Spain.}

\author{Shengjun Yuan}
\affiliation{Key Laboratory of Artificial Micro- and Nano-structures of Ministry of Education 
and School of Physics and Technology, Wuhan University, Wuhan 430072, China}
\affiliation{Wuhan Institute of Quantum Technology, Wuhan 430206, China}

\date{\today}

\begin{abstract}
The recent observed anomalous Hall effect in magic angle twisted bilayer graphene (TBG) aligned to hexagonal boron nitride (hBN) and unconventional ferroelectricity in Bernal bilayer graphene sandwiched by hBN present a new platform to tune the correlated properties in graphene systems. In these graphene-based moir\'e superlattices, the aligned hBN substrate plays an important role. In this paper, we analyze the effects of hBN substrate on the band structure of the TBG. By means of an atomistic tight-binding model we calculate the electronic properties of TBG suspended and encapsulated with hBN. Interestingly, we found that the physical properties of TBG are extremely sensitive to the presence of hBN and they may be completely different if TBG is suspended or encapsulated. We quantify these differences by analysing their electronic properties, optical conductivity and band topology. We found that the narrow bandwidth, band gap, local density of states and optical conductivity are significantly modified by the aligned hBN substrates. Interestingly, these electronic properties can be used as a signature of the alignment in experiment. Moreover, the TBG/hBN superlattices in the presence or absence of the two-fold rotation symmetry response differently to the external electric field. For the TBG suspended in the hBN, application of an electric field results in the charge unevenly distributed between graphene layers, which can be used to tune the strength of the valley Hall effect or the anomalous Hall effect. Such rich topological phase diagram in these systems may be useful for experiments. 

\end{abstract}

\maketitle

\section{Introduction}
Since the discovery of unconventional superconductivity and correlation effects in twisted bilayer graphene (TBG) near the magic angle by Cao \emph{et al.}~\cite{cao2018correlated,cao2018unconventional}, the so-called field of \textquotedblleft twistronics" has become of great interest to the condensed matter community~\cite{carr2017twistronics}.
In this magic angle at approximately 1.1$^\circ$~\cite{bistritzer2011moire}, the system possesses flat bands near charge neutrality, which are the responsible for most of these exotic behaviours.  
Interestingly, not only two graphene layers have been twisted, but also transition metal dichalcogenides~\cite{zhang2020flat,zhan2020tunability}, monochalcogenides~\cite{kennes2020one}, hexagonal boron nitride (hBN)~\cite{woods2021charge,walet2021flat} and black phosphorus~\cite{cao2016gate}, among others. 
Similar to the TBG, flat bands are observed in most of these twisted two-dimensional (2D) materials.

Experimentally, the devices are usually supported on a substrate which is typically a thin sample of hBN~\cite{wang2019new}. Although hBN has a large gap and has been thought to have a small impact on the electronic properties of the materials which are supporting, it is not the case. As we have shown recently~\cite{long2021accurate} (see also~\citep{Shi2021Conm, Cea2020TBGhbN, Shin2021asy, Mao2021Quasi}), hBN has an important effect on the electronic properties of TBG, when the sample is either supported or encapsulated. We have found that when the sample is placed on top of hBN a gap opens due to the appearance of a mass term as a consequence of the breaking of $\mathcal{C}_{2}$ symmetry~\cite{long2021accurate,cea2020band}.  Furthermore, a splitting in the bands appears due to layer degeneracy breaking. In fact, hBN affects the electronic properties of TBG even when the angle between TBG and hBN is far from alignment. When TBG is encapsulated between two layers of hBN the layer degeneracy can be recovered for certain angles, while the gap still appears.

In a typical experimental setup, an electric field is usually applied to the samples in order to change their doping which could also modulate the electric structures of the samples~\cite{Chittari2019,mukai2021unconventional}.
Therefore, the study of the electronic and optical properties such as the local density of states (LDOS) and the optical conductivity of TBG in combination with a substrate and when an electric field is applied is very compelling.

In this work, we extend our previous study to investigate the aforementioned properties in both the case of TBG supported and encapsulated between hBN layers. 
In Sec.~\ref{sec:methods} we describe the atomic structure of our system and the methods that we employ to perform our calculations.
Then, in Sec.~\ref{sec:ldos} the layer degeneracy in electronic properties like LDOS and charge density distribution are studied. 
We discuss the optical conductivity of the twisted bilayer graphene-boron nitride superstructure in Sec. \ref{optical}.
In Sec.~\ref{sec:elec-field} we explore the response of these systems to an external field. 
In Sec.~\ref{sec:topology} we study the effect of encapsulating TBG on its band topology.
Finally, we give a summary and discussion of our work.

\section{Atomic structures and numerical methods}\label{sec:methods}
\subsection{The Atomic structures}
In this work, we mainly focus on two structures. The first one is a trilayer system composed of TBG and a hBN layer lying on the bottom (TBG/hBN). The second one is a sandwich-like system where TBG is encapsulated by two hBN layers (hBN/TBG/hBN). TBG and hBN can be stacked in different ways to have commensurate structures~\cite{shin2021electron}. In our case, we start by constructing an unrotated trilayer structure, where graphene and hBN bonds are parallel and the center of the supercell has an AAA stacking with a carbon site of graphene and a nitrogen (N) site of hBN share the same in-plane position, $(x,y)=(0,0)$. Then we rotate the top layer graphene ($G_{top}$) and bottom layer hBN ($hBN_{bot}$) with angles $\theta_{tbg}$ and $\theta_{bot}$ with respect to the bottom layer graphene ($G_{bot}$), respectively. In our notation, positive angles correspond to counterclockwise rotations. 

We define the lattice vectors of a hexagonal lattice as $a_1 = a(\sqrt{3}/2,1/2)$ and $a_2 = a(\sqrt{3}/2,{\color{red}-}1/2)$ being $a$ the lattice constant. For graphene $a_{g}=0.246$ nm, and for hBN $a_{hBN}=0.2503$ nm. The lattice mismatch between graphene and hBN is $\delta\approx 1.8$\%. The twist angle of TBG can be solely determined by a coprime integer pair $(m,n)$
\begin{equation}
    \theta_{tbg}=2\arcsin \frac{(m-n)}{2\sqrt{m^2+mn+n^2}}, 
\label{equa:twistmoire}
\end{equation}
with moir\'{e} length
\begin{equation}
    L_{tbg}=\frac{a_g(n-m)}{2|\sin{(\theta_{tbg}/2)}|}=a_g\sqrt{m^2+mn+n^2}.
\end{equation}
The moir\'{e} length of the $G_{bot}$/$hBN_{bot}$ superlattice is~\cite{wang2019new}
\begin{equation}
    L_{hBN}=\frac{(1+\delta)a_g}{\sqrt{\delta^2+2(1+\delta)(1-\cos{\theta_{bot}})}}.
\end{equation}
A commensurate structure of TBG/hBN is obtained when
\begin{equation}
    L_{tbg/hBN}=L_{tbg}=qL_{hBN}
\end{equation}
where $q$ is an integer. 
We focus on a system where $\theta_{tbg}=1.05^\circ$ and $\theta_{bot} = 0.53^\circ$.  For this angle combination, the periodicity of the moir\'e pattern of TBG is identical to that constructed by hBN and $G_{bot}$, and therefore a single moir\'e unit cell can be defined for the combined system~\cite{cea2020band,Zhang2019b}. That is, the three moir\'e lengths have the same value $L_{tbg/hBN}=L_{tbg}=L_{hBN} = 13.4$ nm, and $q=1$. To construct the hBN/TBG/hBN structure, we just add a second hBN layer ($hBN_{top}$) on the top of the trilayer structure, the twist angle between $hBN_{top}$ and $G_{bot}$ is  also $\theta_{top} = 0.53^\circ$. It is important to note that, in order to keep the periodicity of these structures, the lattice constant of hBN is slightly modified. In our case, this implies a strain of about $0.14\%$. 
Such a small value of strain will not affect the structural or electronic properties of hBN, which, therefore, will not change the properties of the TBG/hBN systems that we study in this work~\cite{long2021accurate}. The schematics of TBG/hBN and hBN/TBG/hBN are shown in Fig.~\ref{fig:trilayer}(a) and Fig.~\ref{fig:sandwich}(a), respectively.

\subsection{The tight-binding model}
We adopt a combination of semi-classical molecular dynamics and a tight-binding (TB) model to investigate the electronic properties of TBG/hBN and hBN/TBG/hBN structures. 
After constructing a commensurate supercell, we use semi-classical molecular dynamics, which is implemented in LAMMPS~\cite{plimpton1995fast}, to fully (both in-plane and out-of-plane) relax the graphene layers in the two systems. For intralayer interaction between graphene, we use the reactive empirical bond order potential (REBO)~\cite{brenner2002second}. For interlayer interaction between graphene, we use the  registry-dependent Kolmogorov-Crespi (RDKC) potential developed for 
graphite~\cite{kolmogorov2005registry}. 
We use the same RDKC potential for the C-B and C-N interlayer interaction, but with different strength. 
The interaction strength of C-B and C-N are 60\% and 200\% with respect to the original C-C interaction, respectively~\cite{slotman2015effect}. The hBN layers are fixed in a flat configuration to mimic a bulk or a few layers substrate. We assume that the relaxed structures keep the periodicity of the rigid cases.

The full TB Hamiltonian for graphene and hBN heterostructure can be written as~\cite{long2021accurate}
\begin{align}
\label{ham}
\hat{H}=&-\sum_{i,j}t(\mathbf{R}_i-\mathbf{R}_j)\ket{\mathbf{R}_i}\bra{\mathbf{R}_j}+\sum_i\epsilon(\mathbf{R}_i)\ket{\mathbf{R}_i}\bra{\mathbf{R}_i} \\\nonumber
& +\sum_i V_D(\mathbf{R}_i)\ket{\mathbf{R}_i}\bra{\mathbf{R}_i}, 
\end{align}
where $\mathbf{R}_i$ and $\ket{\mathbf{R}_i}$ represent the atom position and the atomic state at site $i$, respectively,  
$t(\mathbf{R}_i-\mathbf{R}_j)$ is the transfer integral between the atomic states at sites $i$ and $j$, 
$\epsilon(\mathbf{R}_i)$ encodes the carbon, boron and nitrogen onsite energies and $V_D(\mathbf{R_i})$ is the deformation potential resulting from the structural relaxation. For the onsite energy of boron, nitrogen and carbon atoms, we assume~\cite{moon2014electronic}:
\begin{gather}
    \epsilon_B=3.34\; \text{eV},\;\epsilon_N=-1.40\;\text{eV},\;\epsilon_C = 0\;\text{eV}
\end{gather}
The lattice deformation leads to the emergence of periodic scalar and gauge potentials~\cite{San-Jose2014,Jung2017,Sachs2011,Kindermann2012,San-Jose2014a,Wallbank2013,2019-xianqing-effective}. All these effects can be accurately considered in our TB model. To incorporate the relaxation effect into Hamiltonian in Eq. (\ref{ham}), we introduce the deformation potential term as~\cite{slotman2015effect}:
\begin{equation}
    V_D(\mathbf{R}_i)=g_1 \frac{S(\mathbf{R}_i)-S_0}{S_0},
\end{equation}
where the screened deformation potential $g_1=4$ eV~\cite{ochoa2011temperature}, $S(\bm{R_i})$ is the effective area of site $i$ that is modulated by local deformations, and $S_0=\sqrt{3}a/4$ is the effective area in equilibrium. 
For the transfer integral, we simply adopt the common Slater-Koster-type function for any combination of atomic species~\cite{moon2014electronic}:
\begin{gather}
    -t(\mathbf{R})=V_{pp\pi}[1-\left(\frac{\mathbf{R}\cdot \mathbf{e}_z}{R}\right)^2]+V_{pp\sigma}\left(\frac{\mathbf{R}\cdot \mathbf{e}_z}{R}\right)^2, 
\end{gather}
where
\begin{gather}
    V_{pp\pi}=V_{pp\pi}^0 e^{-\frac{R-a_0}{r_0}}, \\
    V_{pp\sigma}=V_{pp\sigma}^0 e^{-\frac{R-d_0}{r_0}}.
    \label{eq: vpisigma}
\end{gather}
In the above equation $\mathbf{e}_z$ is the unit vector perpendicular to the graphene plane, $R = |\mathbf{R}|$, $a_0=a_g/\sqrt{3} \approx 0.142 $ nm is the C-C distance, $d_0=0.335$ nm is the interlayer distance, $V_{pp\pi}^0$ and $V_{pp\sigma}^0$  are the intralayer and interlayer transfer integrals between nearest neighbor atoms, respectively. We take $V_{pp\pi}^0 \approx -2.7$ eV, $V_{pp\sigma}^0 \approx 0.48$ eV. The parameter $r_0$ is the decay length of the transfer integral, and is chosen as $0.184a_g$ so that the next nearest intralayer coupling becomes  $0.1V_{pp\pi}^0$. For atoms whose distance is more than $0.6$ nm, we set $t(\mathbf{R})=0$ since for larger distances the value of hopping energy is small enough to be safely neglected.

\subsection{The electronic properties}
Once the TB Hamiltonian is constructed, we can calculate the electronic properties of the TBG/hBN superlattices. Since the TBG/hBN and hBN/TBG/hBN structures contain tens of thousands of atoms, we use a tight-binding propagation method (TBPM) to obtain the density of states (DOS) and optical conductivity. The TBPM is based on the numerical solution of the time-dependent Schr\"odinger equation and requires no diagonalization processes, which is implemented in our home-made TBPLaS simulator~\cite{yuan2010modeling,li2022tbplas,tbplas}.  
In TBPM, a random initial state $|\phi_0\rangle$ is used with $\langle \phi_0| \phi_0\rangle =1$. The density of states is calculated as a Fourier transform of the time-dependent correlation function
\begin{equation}
    D(E) = \frac{1}{2\pi}\displaystyle \int_{-\infty}^{+\infty}e^{iE\tau}\langle\phi_0|e^{-iH\tau/\hbar}|\phi_0\rangle d\tau
\end{equation}
The optical conductivity is calculated by combining the Kubo formula with TBPM~\cite{yuan2010modeling}. The real part of the optical conductivity matrix, $\sigma_{\alpha\beta}$, at temperature $T$ reads
\begin{equation}
\begin{split}
    \Re\;\left\lbrace\sigma_{\alpha\beta}(\omega)\right\rbrace =& \lim_{E \to 0^+} \frac{e^{-\hbar \omega/k_B T}-1}{\hbar \omega A} \int_0^{\infty} e^{-E \tau}\sin\left(\omega\tau\right) \\
    &\times 2 \Im\left\lbrace\langle \phi_2(\tau)|j_\alpha| \phi_1(\tau)\rangle_\beta\right\rbrace \mathrm d\tau,
\end{split}
\label{op}
\end{equation}
with $A$ the area of the unit cell, and $|\phi_1(\tau)\rangle$ and $|\phi_2(\tau)\rangle$ read
\begin{equation}
    \begin{split}
        |\phi_1(\tau)\rangle_\alpha &= e^{-iH\tau/\hbar}[1-f(H)]J_\alpha|\phi_0\rangle, \\
        |\phi_2(\tau)\rangle &= e^{-iH\tau/\hbar}f(H)|\phi_0\rangle, \\
    \end{split}
\end{equation}
where $f(H) = 1/(e^{(H-\mu)/k_B T}+1)$ is the Fermi-Dirac distribution operator and $\mu$ is the electronic chemical potential. In this work, all the optical conductivity are calculated at $T=300$ K and $\mu=0$. In the TBPM, the convergence can be guaranteed by averaging over different initial states $|\phi_0\rangle$. For large enough systems, the results are converged with only one random initial state.
The LDOS is obtained via the recursion method in real space based on the Lanczos algorithm~\cite{haydock1972electronic}. The eigenstates and eigenvalues are obtained by direct diagonalization of the Hamiltonian in Eq. (\ref{ham}).

\begin{figure*}[t!]
\centering
\includegraphics[width=0.9\textwidth]{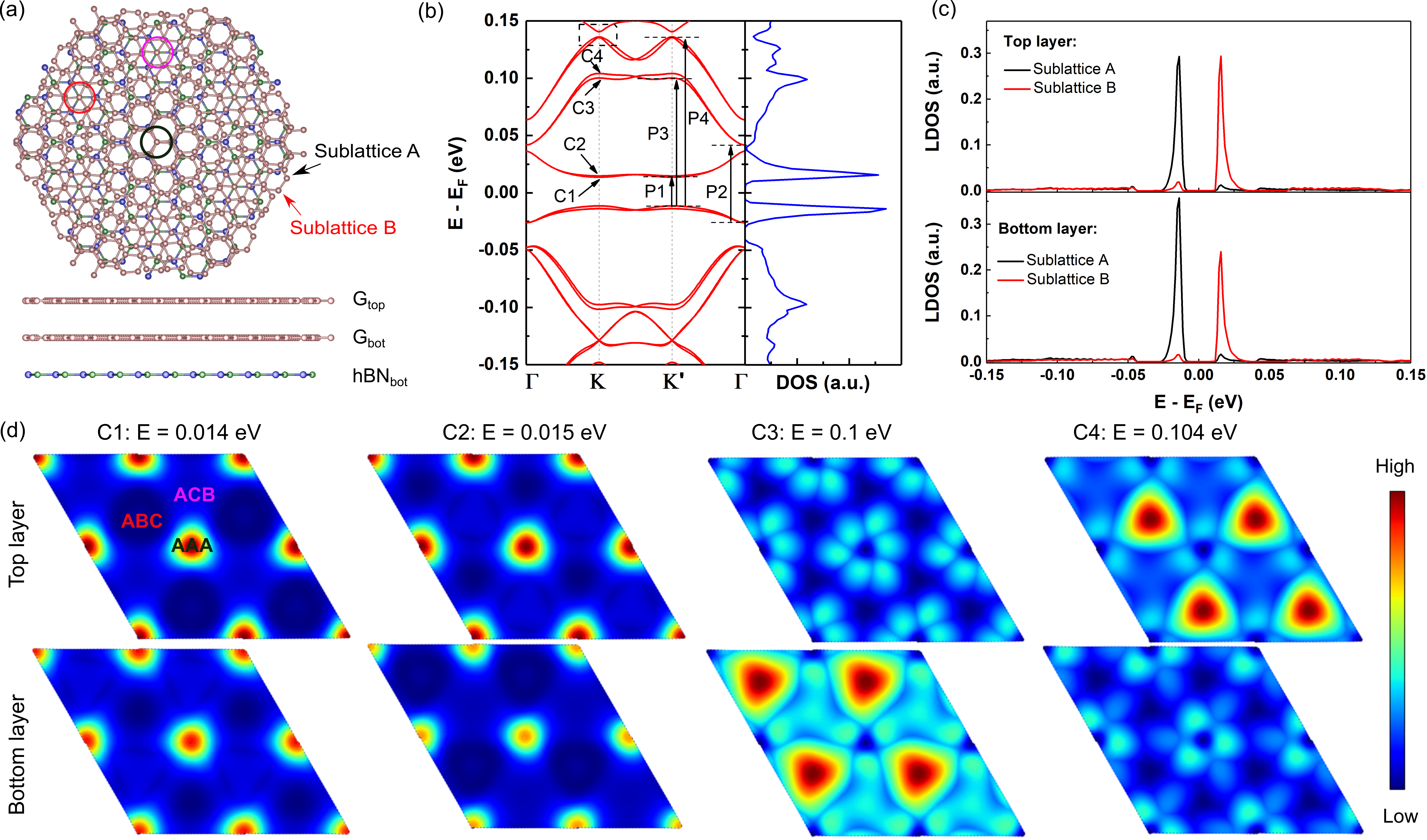}
\caption{(a) The top (upper panel) and side (lower panel) views of the trilayer structure. The center of the top view is the AAA high-symmetry stacking, which has the carbon atoms of graphene and nitrogen atom of hBN share the same in-plane position. The sublattices A and B of the graphene are specified. The high-symmetry stackings of AAA, ABC and ACB are outlined with black, red and purple circles, respectively. (b) Band structure and density of states of the trilayer structure. The black arrows indicate various significant optical transitions. (c) Local density of states of sublattices A and B in the AAA stacking region. The LDOS of atoms from top (upper panel) and bottom (lower panel) graphene layers are plotted separately. (d) Eigenstates $|\psi|^2$ in real space. The eigenstates are calculated by diagonalizing the Hamiltonian in Eq. (\ref{ham}). The corresponding energies  $C1$, $C2$ and $C3$ of these eigenstates are illustrated in (b).}
\label{fig:trilayer}
\end{figure*}

\begin{figure*}[t!]
\centering
\includegraphics[width=0.9\textwidth]{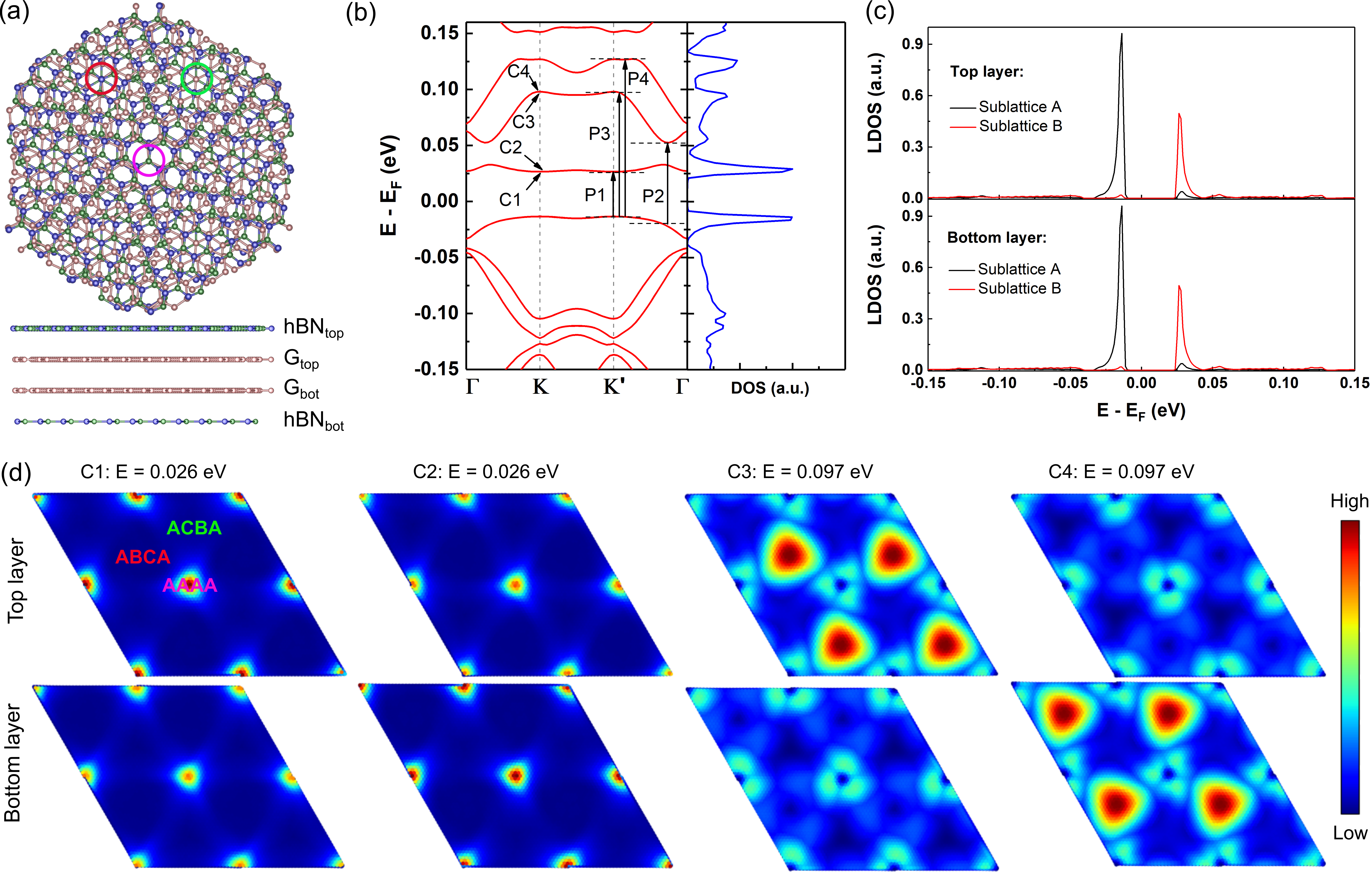}
\caption{(a) The top (upper panel) and side (lower panel) views of the tetralayer structure. The center of the top view is the AAAA high-symmetry stacking, which has the carbon atoms of graphene and nitrogen atoms of hBN share the same in-plane position. The high-symmetry stackings of AAAA, ABCA and ACBA are outlined by purple, red and green circles, respectively. (b) Band structure and density of states of tetralayer structure. The black arrows indicate various significant optical transitions. (c) Local density of states of sites in the AAAA stacking region. (d) Eigenstates $|\psi|^2$ in real space. The corresponding energies, $C1$, $C2$ and $C3$ of these eigenstates are illustrated in (b).}
\label{fig:sandwich}
\end{figure*}

\section{TBG Heterostructures}\label{sec:ldos}

\subsection{TBG supported on hBN}

In free standing TBG, the layer degree of freedom is disentangled from spin and valley, forming eight-fold degeneracy in the low energy flat bands. As shown in Fig. \ref{fig:tbg}, the conduction and valence flat bands are connected by Dirac points at the K and K' points of the moir\'e Brillouin zones (mBZ). These Dirac points are protected by $\mathcal{C}_2\mathcal{T}$ symmetry, where $\mathcal{C}_2$ is the two-fold rotation operator and $\mathcal{T}$ is the time\textendash reversal operator. If $\mathcal{C}_2\mathcal{T}$ is broken, the Dirac points will become gapped. As we discussed in our previous work~\cite{long2021accurate}, in a TBG/hBN superlattice, the hBN substrate introduces two contributions to the TBG system: the first one is the energy difference between nitrogen and boron atoms. This gives rise to different adhesion energies in the TBG/hBN system and breaks the $\mathcal{C}_2$ inversion symmetry, which is the main source of this mass gap. The second contribution is the lattice relaxation effects that give rise to a deformation potential and a pseudo-magnetic field~\cite{Shi2020Nature}. Such relaxation effect ensures the persistence of a gap opening in the Dirac point for angles between TBG and hBN far from alignment~\cite{jung2015origin}.

The substrate effects can be clearly observed in the band structure shown in Fig. \ref{fig:trilayer}(b). 
Narrow bands are separated by a gap of around 30 meV due to the breaking of $\mathcal C_2$ symmetry~\cite{jung2015origin,Hunt2013,SL13,Amet2013,Getal14,Cetal14,Yankowitz2014,Wetal15,Lee_science2016,Wetal16,Yankowitz_nat2018,Kim2018}. This first gap (energy difference between the flat bands at K) is reduced when the angle $\theta_{bot}$ increased~\cite{long2021accurate}. 
Moreover, the presence of a substrate acting on a single graphene layer breaks the mirror symmetry between layers and their Dirac cones are shifted in energy. This layer degeneracy breaking is responsible for the observed splitting between narrow bands. This splitting is more obvious in the remote bands located at around $\pm 0.1$ eV. As we will discuss later, a perpendicular electric field will further increase this splitting. 
The narrow bands show a significant electron-hole asymmetry due to the strong superlattice potential induced by the hBN. Compared to TBG (see Fig.~\ref{fig:tbg}) the flat bands become dispersive and the peaks in the DOS of the TBG at charge neutrality are smoothed giving rise to an insulating structure in Fig. \ref{fig:trilayer}(b).  
In graphene/hBN superlattices, secondary Dirac cones appear at higher energies, which can be attributed to the moir\'e potential~\cite{slotman2015effect}. 
In the TBG/hBN structure that we study, the periodicity of the moir\'e pattern is 14 nm. This entails the secondary Dirac cones to be located at around $0.14$ eV (outlined by a dashed rectangle in Fig. \ref{fig:trilayer}(b)). The states of the induced Dirac cones are mainly localized at the ABC or ACB stacking regions (the results not shown here), which is similar to the results of graphene/hBN superlattices~\cite{slotman2015effect}.

The hBN substrate also significantly modifies the local properties of the TBG/hBN superlattice. 
In free standing TBG (see Fig. \ref{fig:tbg}), the flat band states are mainly localized in both sublattices A and B around the AA stacking regions. If we look at the calculated LDOS of sublattices A and B at the AAA stacking region, as shown in Fig. \ref{fig:trilayer}(c), the peak in the sublattice B is lower than that of the sublattice A in the bottom layer. 
This is due to the role that hBN is playing as a substrate. The hBN substrate breaks the sublattice symmetry in the $G_{bot}$, making the LDOS peaks in the sublattice B lower than the sublattice A. On the contrary, the hBN substrate has negligible influence on the top layer. The states from $G_{top}$ and $G_{bot}$ contribute unequally to the conduction band and valence band, which is the natural result of the breaking of the layer degeneracy. In some works, the hBN substrate effect on the TBG is introduced via an effective periodic potential acting on the nearest graphene layer~\cite{moon2014electronic,San-Jose2014}. 
Our results indicate again that this approach is also correct. We also plot the real space wave function of the narrow bands. 
Different from the TBG case shown in Fig.~\ref{fig:tbg}, where states in the narrow bands are localized around the AA centers. In the presence of hBN we observe some states localized in the ABC or ACB stacking regions. That is, a small part of states from the C1 band are localized in the ACB region (see labels in Fig. \ref{fig:trilayer}(b)), whereas some states from the C2 band are in the ABC region. Such difference is a consequence of the substrate potential that redistribute the charges within the moir\'e unit cell. The states from a higher energy of C3 are mainly localized in the ACB region of the bottom layer. 

\subsection{TBG encapsulated in hBN}

We now consider a tetralayer structure where, as shown in Fig. \ref{fig:sandwich}(a), TBG is encapsulated by two hBN layers. As described in our previous work~\cite{long2021accurate} there are several possible stacking configurations for the tetralayer structure (see also~\cite{shin2021electron}). In this work  we choose the twist angle and stacking configuration such that the mirror symmetry between layers is recovered, this  is given by the condition $|\theta_{bot}|=|\theta_{top}|$.  Therefore, we could tune the layer degeneracy by adding or removing one of the hBN layers (as we will describe in the following section, an electric field can also be used to tune the layer degeneracy). It is important to mention that the breaking of the layer degeneracy is because the Dirac cones in each graphene layer are being affected by their nearest hBN layer. A direct consequence of recovering the layer degeneracy is the disappearance of the splitting in the flat bands, as we can see from the comparison between the band structure in Fig.~\ref{fig:trilayer}(b) with that  in Fig.~\ref{fig:sandwich}(b). The splitting does not only disappear in the narrow bands near the Fermi level, but also in the high energy bands, indicating an intrinsic change in the electronic structure of the encapsulated structure. Compared with TBG, both middle bands become narrower. The gap between the lower narrow bands has a value of around 50 meV. Moreover, as we can see from the LDOS calculation, Fig. \ref{fig:sandwich}(c), the states from $G_{top}$ and $G_{bot}$ contribute equally to both narrow bands. Another noteworthy result is that the states of both bands have different contributions from different sublattices. That is, both the top and bottom layers have a different sublattice charge distribution, with the LDOS peaks of the sublattice B lower than the sublattice A. Regarding the charge density maps in Fig.~\ref{fig:sandwich}(d) and compared to the TBG/hBN superlattice, the states of the narrow bands become more localized in the AAAA regions, forming a triangular shape with smaller area. Moreover, the states of the two upper narrow bands have a quite similar localization in real space.

\section{Optical conductivity}
\label{optical}
\begin{figure}[t!]
\centering
\includegraphics[width=0.48\textwidth]{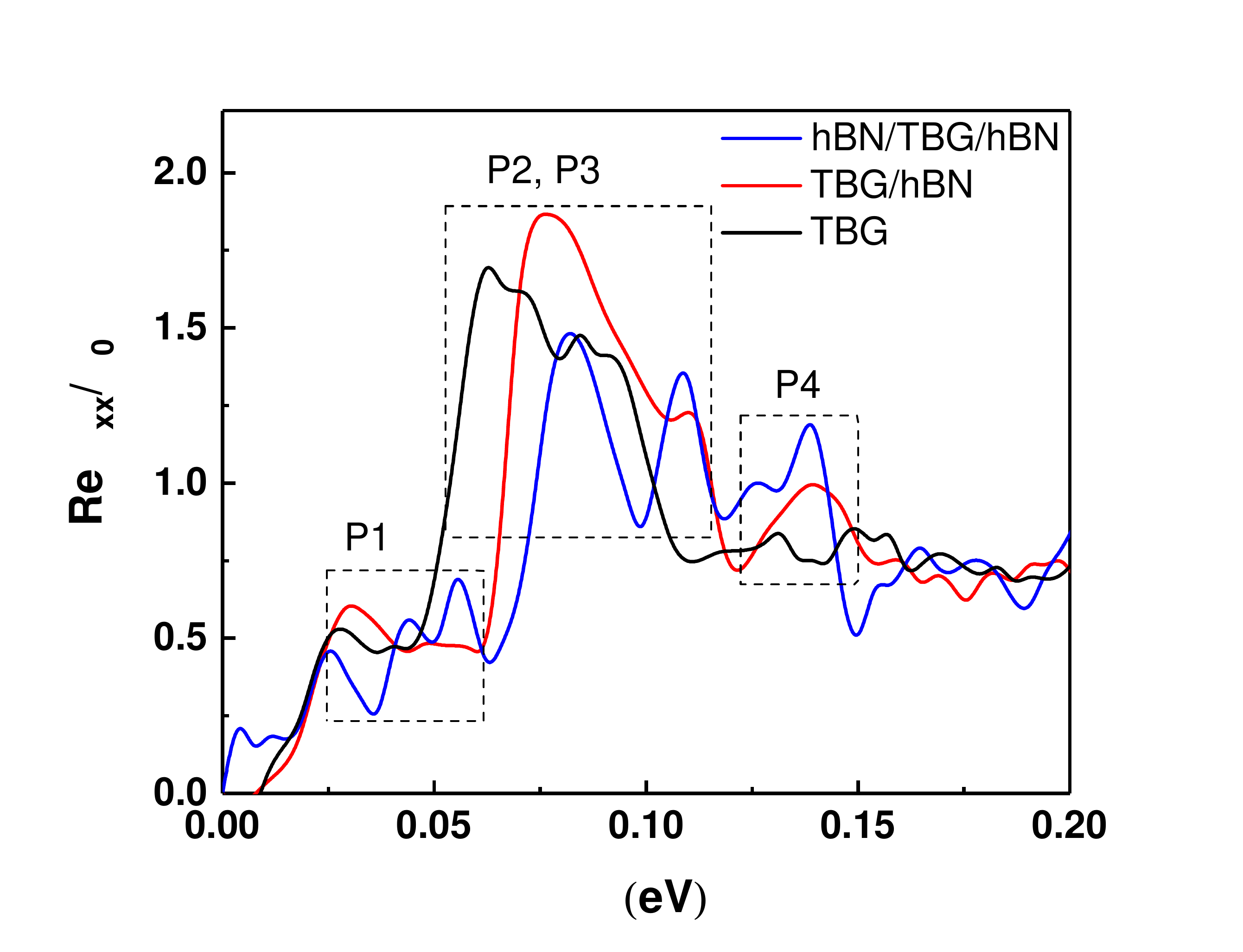}
\caption{Calculated optical conductivity of TBG (black line), TBG/hBN (red line) and hBN/TBG/hBN (blue line). The chemical potential and temperature are $\mu=0$ and $T=300$ K, respectively. The optical conductivity has been normalized to the universal optical conductivity of graphene $\sigma_0=\pi e^2/2h$. Significant optical peaks P1, P2, P3 and P4 are illustrated by a dashed rectangle.}
\label{fig:Figure3opcond}
\end{figure}

In the previous sections, we have shown that the hBN strongly modified the electronic structure of TBG. In this part, we discuss how will the substrate modify the optical conductivity, which can be conveniently explored, for example, by means of infrared and terahertz spectroscopies~\cite{ni2015plasmons}. 
The real part of the longitudinal optical conductivity of different TBG moir\'e systems computed from Eq.~(\ref{op}) are shown in Fig.~\ref{fig:Figure3opcond}. In a graphene/hBN superlattice, the moir\'e period leads to the emergence of minibands around the extra Dirac point and the main signatures of the optical excitations are found between the valence and conduction minibands~\cite{slotman2015effect}. 
In TBG/hBN there are multiple transitions, as shown in Fig.~\ref{fig:Figure3opcond}. 
Black line is the optical conductivity of pristine TBG. 
Looking at the optical conductivity of pristine TBG we assign the peak around 0.03 eV (P1) to the transition between middle narrow bands. 
This peak is shifted to higher energies when TBG is supported (red line) or encapsulated (blue line) with hBN.
For the encapsulated case, the shift is larger.
Peaks P2/P3 and P4 around 0.075 eV and 0.15 eV, respectively, correspond to transitions between the narrow and high energy bands. 
Peaks P2/P3 have the same behaviour as peak P1 when TBG is either supported or encapsulated.
These energy shifts in the optical conductivity may allow to identify the degree of alignment of hBN with TBG. In an experiment, if two different regions or two samples have an hBN substrate they may give different signals in the optical conductivity. If the energy difference of the maximum peaks is large as in Fig.~\ref{fig:Figure3opcond} then the hBN is modifying the band structure and this may be a signature of a near alignment situation. If the difference between signals is negligible, then the hBN substrate is unaligned and is not affecting the TBG bands~\cite{long2021accurate}. 

\begin{figure*}[t!]
\centering
\includegraphics[width=\textwidth]{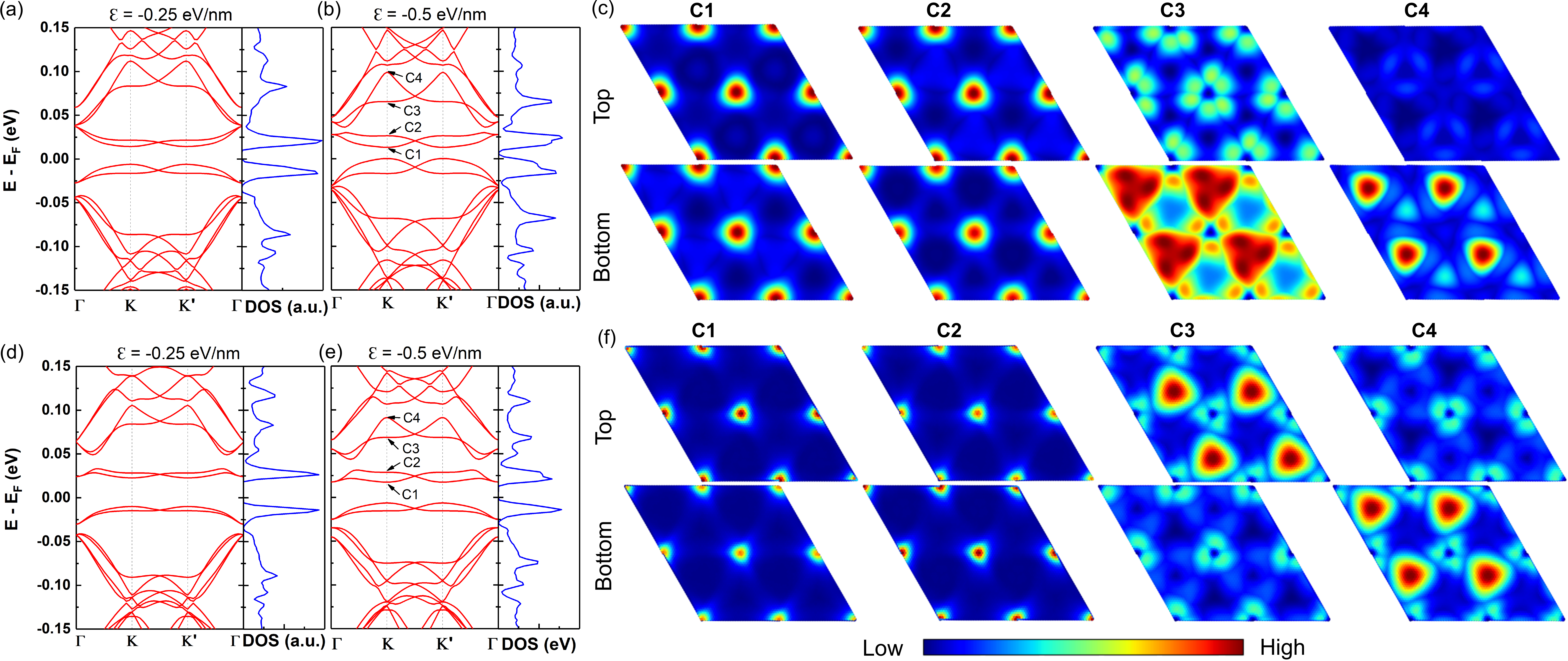}
\caption{(a)(b)(d)(e) Band structures and density of states for trilayer structure and tetralayer structure with different electric field, respectively, (c) and (f) Eigenstates $|\psi|^2$ that calculated from diagonalization of Hamiltonian for trilayer and tetralayer structure, respectively. The corresponding energies of these eigenstates are illustrated in (b) and (e). In the density plots, red is for the maximum and dark blue for the minimum charge density.}
\label{fig:electric}
\end{figure*}

\section{The effect of a perpendicular electric field} \label{sec:elec-field}
It has been shown that a perpendicular electric field in graphene/hBN superlattices allows the tunability of their physical properties~\citep{pantaleon2021narrow}, such as the case of Bernal bilayer graphene on hBN (BG/hBN)~\cite{moriyama2019observation}, ABC stacked graphene on hBN (TG/hBN)~\cite{chen2019signatures} or twisted double bilayer graphene (TDBG)~\cite{he2021symmetry}, among others. Interestingly, the bandwidth of the narrow bands in the first three systems is reduced by increasing the electric field~\cite{pantaleon2021narrow,he2021symmetry,carr2020ultraheavy} while in the latter, the bandwidth is increased~\cite{Lei2021EFieldTTG,lopez2020electrical}. 
In TBG a perpendicular electric field shifts the degeneracy of the Dirac cones~\cite{Koshin02020EField,PabloSJ2013HelEfield,Koshino2014Efield} and does not open a gap between the narrow bands. The cones are protected by $\mathcal{C}_{2} \mathcal{T}$ which is preserved even in the presence of the electric field~\cite{Koshino2014Efield,Po2018}. The presence of gap between narrow bands in TBG with hBN is because the substrate breaks $\mathcal{C}_{2}$. To introduce an electric field in our TB model, we add an on-site potential with different sign on each graphene layer. The electrostatic potential is given by $\Delta V= d_0 \mathcal{E}\cdot\mathbf{e}_z$ being $\mathcal{E}$ the vertical electric field which has negative value if its direction is opposite to $\mathbf{e}_z$. Then, an extra onsite term is added in the Hamiltonian in Eq. (\ref{ham}) as
\begin{equation}
H_V = \pm \frac{\Delta V}{2}\ket{\boldsymbol{R_i}}\bra{\boldsymbol{R_i}}.
\label{eq: VertField}
\end{equation}
Here +(-) corresponds to $G_{bot}$($G_{top}$). 

In pristine TBG, mirror symmetry ensures the degeneracy of the Dirac cones. In the presence of the electric field this symmetry is broken because at each layer an on-site potential with different sign is introduced and this is the main source of the gap between narrow bands (see Ref.~\cite{long2021accurate} for further details). 
Indeed, this effect can be clearly seen in Fig.~\ref{fig:electric}.
In the TBG/hBN system (top row of Fig.~\ref{fig:electric}) the degeneracy of the Dirac cones (or layer degeneracy) is broken because the hBN is acting on a single graphene layer. 
Notice that the energy bands corresponding to each valley are shifted in opposite directions. The effect of the electric field is a further shifting of the energy bands and this effectively looks like a gap closing as shown in the DOS in Fig.~\ref{fig:electric}(b) and (e). 
For the encapsulated system (bottom row of Fig.~\ref{fig:electric}) we can see that the electric field shifts the energy bands, however, the effect of the field is less drastic because the layer degeneracy is preserved (see Fig.~\ref{fig:sandwich}(b)). 
The effects of the layer degeneracy breaking can be observed in the density maps in Fig.~\ref{fig:electric}(c) and (f). As expected in TBG systems, the charge is strongly localized around the AA centers even with an hBN substrate. In the presence of both hBN and an electric field, the charge is unevenly distributed between layers inducing a polarization. The latter is stronger in the suspended situation, as shown in Fig.~\ref{fig:electric}(c). We have found that the layer polarization is a physical phenomena resulting from the layer degeneracy breaking. While this degeneracy can be recovered in TBG by encapsulation with hBN, c.f Fig.~\ref{fig:sandwich}, with an electric field is always broken. Our results indicate that in TBG samples suspended or encapsulated with hBN, the presence of an electric field polarizes the charges within the layers. From time\textendash reversal symmetry, the band energy shift is opposite in each valley and therefore the charge polarization is a valley phenomena. The total charge distribution in each layer is the same if both valleys are considered. This effect can be used to tune the strength of the valley Hall effect or the anomalous Hall effect in the presence of a magnetic field.

\section{Stacking dependent band topology}\label{sec:topology}

\begin{figure}
\centering
\includegraphics[width=0.45\textwidth]{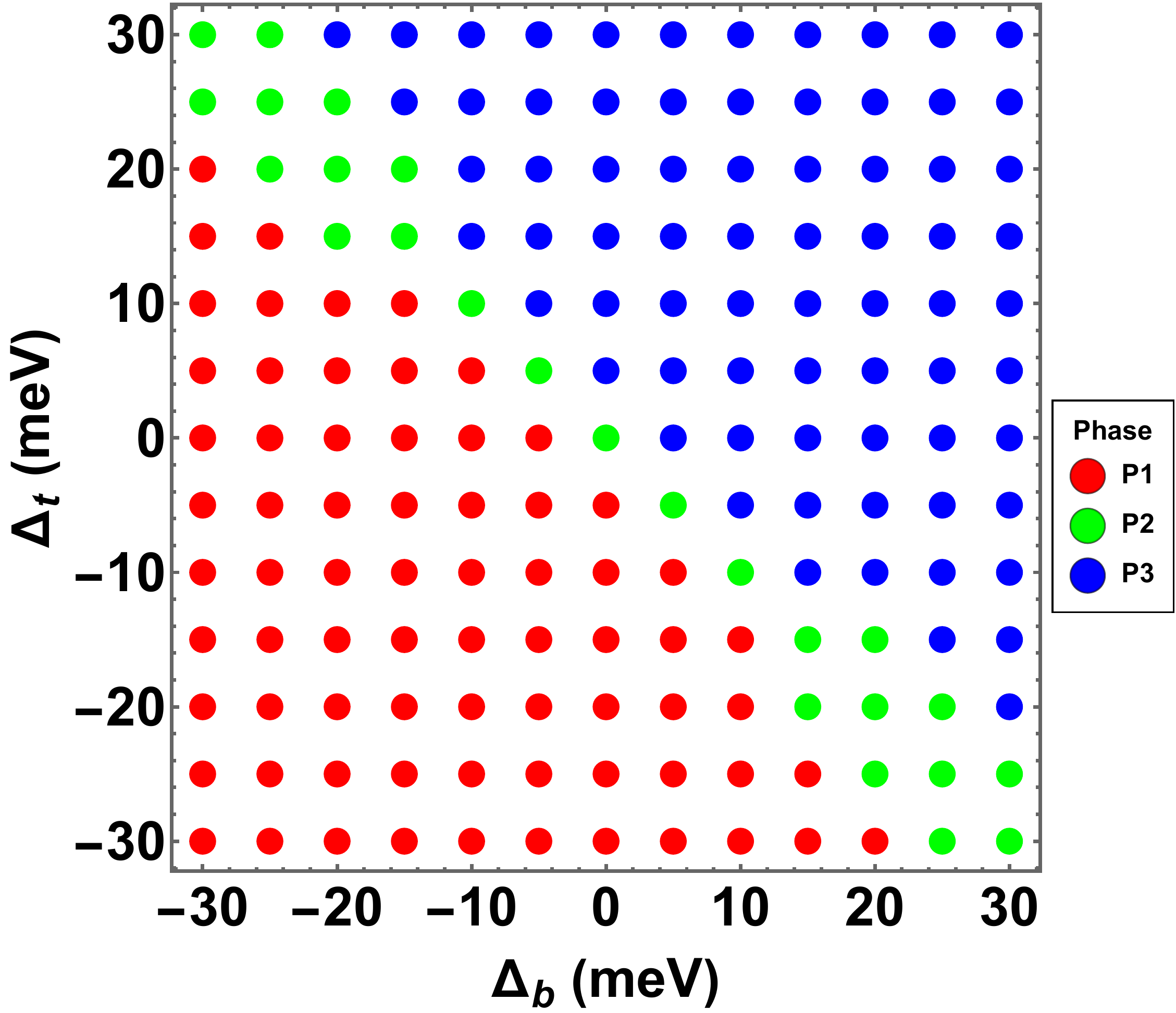}
\caption{Topological phases of the two narrow middle bands of TBG with hBN as a funtion of the mass terms. We distinguish three topological phases, $\{\mathcal{C}_{b}, \mathcal{C}_{t}\}$, identified as $P_1 = \{-1, 1\}$, $P_2 = \{0, 0\}$ and $P_3 =  \{1,-1\}$}
\label{fig:mass}
\end{figure}

Theoretical studies analysing the topology and correlated effects in the presence of a substrate usually introduce the effect of the hBN by considering only the mass term~\cite{Zhang2019b, Repellin2020Frac, Bultinck2020Fer}. While this approach is valid, in experiments the TBG samples are supported on~\citep{Serlin2019} or encapsulated~\cite{Setal20,Ma2020Enc} with hBN and this may result in different band topology. In this section we show that the topological properties of TBG/hBN and hBN/TBG/hBN can be completely different. In both systems, the breaking of inversion symmetry in the TB model allows for a non-zero Berry curvature with opposite signs in each valley. Because of time\textendash reversal symmetry, the Berry curvature in each valley has opposite sign and hence the total Chern number of a given band is zero. However, the topological invariants can be defined for each valley~\citep{San-Jose2014a,Song2015}. The finite Berry curvature for a single valley is given by
\begin{equation}
\Omega_{\vec{k},l} = 2~\Im\left\lbrace \braket{\partial_{k_x}\psi_{\vec{k},l}|\partial_{k_y}\psi_{\vec{k},l}}\right\rbrace,
\label{eq:Omega}
\end{equation}
where $l$ is a band index with energy $E_l(\vec k)$, momentum $\vec{k}=\{k_x,k_x \}$ and wavefunctions $\psi_{\vec{k},l}$. For the different stacking configurations, the bands for each valley are isolated and their Berry curvature is well defined. Therefore, we can assign a valley Chern number, ${{\cal C}_{l}}$, to the band $l$ which is given by the integral of the Berry curvature in the moir\'e Brillouin zone
\begin{equation}
\label{eq: Chern}
{\cal C}_{l} = \frac{1}{2 \pi} \int_\mathrm{mBZ} d^2 \vec k \Omega_{\vec{k},l}.
\end{equation}
We use the algorithm in Ref.~\cite{Fukui2005} and the continuum model in Ref.~\cite{long2021accurate} to compute the Berry curvature. To simplify our analysis we are considering only the mass terms, here, $\Delta_{b/t}$ are mass terms acting in the bottom and top graphene layer, respectively, mimicking an hBN substrate, we also consider the bands only at charge neutrality because the band topology may also depend on the filling fraction~\cite{Pantalen2022}. Figure~\ref{fig:mass} displays the set of valley Chern numbers for the systems considered in this work. We can distinguish three topological phases with Chern numbers for each band, $\{\mathcal{C}_{b}, \mathcal{C}_{t}\}$, identified as $P_1 = \{-1, 1\}$, $P_2 = \{0, 0\}$ and $P_3 =  \{1,-1\}$. The phases $P_1$ and $P_3$ are the usual phases found in the presence of a single hBN layer. However, for the encapsulated situation there are different possibilities, such is the case of Fig.~\ref{fig:sandwich}(b) where the corresponding phase is $P3$ because we found that the mass terms have the same sign and magnitude ($\sim 30$ meV). Additional stacking configurations may give different topological phases, for example if we swap the Boron by the Nitrogen in the stack in Fig.~\ref{fig:sandwich}(a) the mass terms change sign and the phase can be $P1$ or $P2$. Interestingly, in the encapsulated situation, if each hBN is nearly aligned with their adjacent TBG  the topological phase is $P2$ if the mass terms have different sign and similar magnitude. This situation can be achieved, for example, aligning both hBN layers with respect to each other or by rotating them in opposite directions with the same angle. The dominant contribution in TBG/hBN structures are the mass terms. If the angle between hBN and its adjacent graphene layer is increased, these terms decrease and may survive up to $3^\circ$ of alignment (see Ref.~\cite{long2021accurate} for further details). Therefore, Fig.~\ref{fig:mass} is a general result and indicates the rich band topology of suspended and encapsulated samples of TBG with hBN that can be obtained near alignment.  

\section{Conclusions}

We have studied the effects of an hexagonal boron nitride substrate in the electronic and topological properties of TBG. In particular, we have studied TBG on hBN and TBG encapsulated with hBN. By using a real space tight-binding method in combination with semi-classical molecular dynamics, we calculate the band structure, DOS, LDOS, optical conductivity and analyze the stacking-dependent topological properties of these systems. We find that the substrate significantly modifies the electronic properties of the TBG due to the broken $\mathcal C_2$ symmetry. Compared to the free standing TBG, the narrow bands have been strongly modified, and separated by a gap of around 30 meV. In the TBG/hBN system we found that the substrate induces a layer degeneracy breaking which results in an splitting of the TB band structure and uneven distribution of the LDOS in a single layer. In the encapsulated TBG/hBN system we found that the layer degeneracy is recovered if the twist angle between each graphene and its nearest hBN layer is of the same magnitude. In addition, we calculate the DOS peaks in both system and we found that the optical peaks corresponding to the different transitions are remarkably different in both energy positions and magnitudes. Such energy shifts in the optical peaks may provide a method to identify if TBG is aligned with the hBN substrate. Both considered heterostructures are also strongly sensitive to a perpendicular electric field. A direct consequence of this field is the shifting of the energy bands which may look as a gap closing in the DOS. Interestingly, by mapping the real space distribution of the wavefunctions we found that the electric field polarizes the charge in each valley which can be used to tune the strength of the valley Hall effect or the anomalous Hall effect in the presence of a magnetic field. Finally, by calculating the valley Chern numbers, we found that depending on the induced mass gap, different topological phases can be obtained. Because the mass gap is a consequence of the degree of the hBN alignment, this result, combined with the optical conductivity peaks may give a simple methodology to identify the encapsulating conditions in TBG/hBN samples.

\section*{ACKNOWLEDGMENTS}
This work was supported by the National Natural Science Foundation of China (Grant No. 11774269) and the National Key R\&D Program of China (Grant No. 2018YFA0305800). 
IMDEA Nanociencia acknowledges support from the ``Severo Ochoa" Programme for Centres of Excellence in R\&D (Grant No. SEV-2016-0686).
P.A.P and F.G. acknowledge funding from the European Commission, within the Graphene Flagship, Core 3, grant number 881603 and from grants NMAT2D (Comunidad de Madrid, Spain), SprQuMat (Ministerio de Ciencia e Innovaci\'on, Spain). 
Numerical calculations presented in this paper have been performed on the supercomputing system in the Supercomputing Center of Wuhan University.

\begin{figure*}[h!]
\centering
\includegraphics[width=0.9\textwidth]{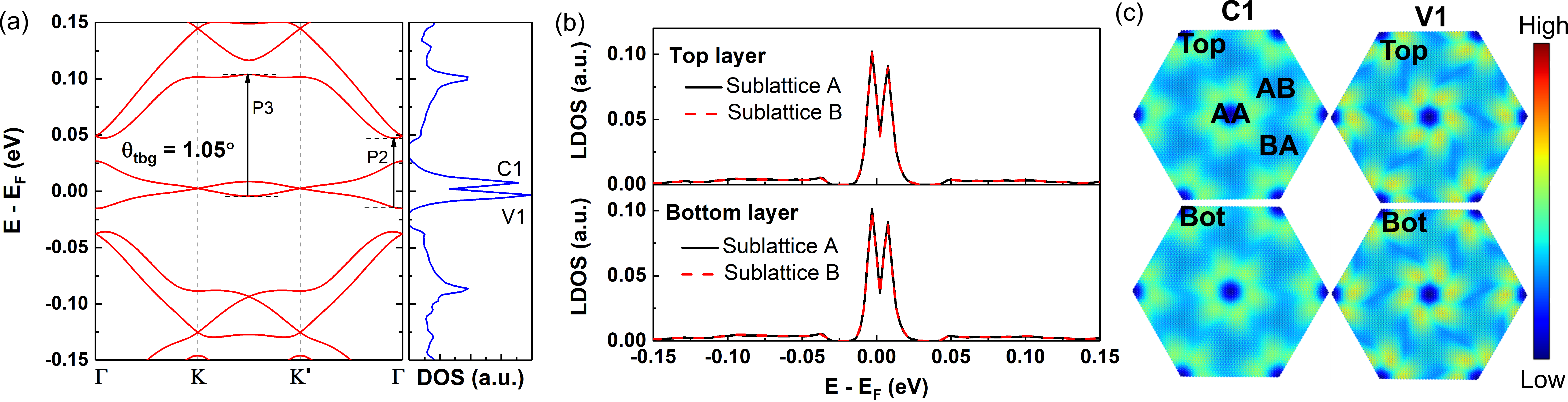}
\caption{(a) Band structure and density of states, (b) LDOS of the sublattice A and B in the AA stacking region and (c) LDOS mapping in real space for free standing twisted bilayer graphene with $\theta=1.05^\circ$. The LDOS mapping are obtained via the TBPM method. The corresponding energies of the eigenstates are illustrated in the DOS.}
\label{fig:tbg}
\end{figure*}
\appendix
\section{The electronic properties of free standing twisted bilayer graphene}\label{app:A}
In this section, we briefly describe the properties of pristine TBG. The hopping parameters in Eq.~(\ref{eq: vpisigma}) are given such that the \textquotedblleft magic" angle is at $1.21^\circ$. Figure~\ref{fig:tbg}(a) show the band structure of TBG  with a twist angle $\theta_{tbg}=1.05^\circ$. The DOS display the van Hove singularities (vHs) due to the two narrow bands. Figure~\ref{fig:tbg}(b) shows that the contribution to each sublattice to the LDOS is identical due to the preserved sublattice, layer and valley symmetries. In Fig.~\ref{fig:tbg}(c) we show the LDOS maps in real space for energies at the vHs. They have the familiar \textquotedblleft fidget-spinner" shape~\cite{Po2018} with the states localized around the AA region.

\bibliographystyle{apsrev4-2}
\bibliography{valley_working_version.bib}

\end{document}